\documentclass[12pt]{article} \textwidth=6in \oddsidemargin=0in
\usepackage{amssymb} 
\begin{document}
\def\ov{\over} \def\be{\begin{equation}} \def\s{\sigma}
\def\ee{\end{equation}} \def\iy{\infty} \def\({\left(} \def\){\right)}  
\def\x{\xi} \def\inv{^{-1}} \def\fh{\hat f} \def\gh{\hat g}
\def\bc{\begin{center}} \def\ec{\end{center}} \def\d{\delta}
\def\R{\mathbb R} \def\sg{{\rm sgn}} \def\G{\Gamma} \def\ph{\varphi} \def\e{\eta} \def\ch{\chi} \def\r{\rho} \def\g{\gamma} \def\fe{\mathfrak{e}}
\def\ep{\varepsilon} \def\ps{\psi} \def\Ph{\Phi} \def\k{\kappa} 
\def\ap{\approx} \def\z{\zeta} \def\iyy{_{-\iy}^\iy} \def\ziy{_0^\iy}
\def\arg{{\rm arg}\,} \def\sgn{{\rm sgn}\,} \def\Re{{\mathcal Re}\,} 
\def\Ft{Fourier transform\ }  \def\ch{\raisebox{.4ex}{$\chi$}} 
\newcommand{\twotwo}[4]{\left(\begin{array}{cc}#1&#2\\&\\#3&#4\end{array}\right)} \def\pmp{{\pm p}} \def\fe{\mathfrak{e}}
\newcommand {\twoone}[2]{\left(\begin{array}{c}#1\\ \\#2\end{array}\right)}
\def\C{\mathcal C} \def\irr{_{-r/2}^{r/2}}  \def\wh{\widehat}
\def\F{\mathcal F} \def\CC{\mathbb C} \def\L1h{\wh{L^1}} \def\I{{\rm Im}\,}
\def\LL{\mathcal L} \def\noi{\noindent} \def\sp{\vspace{1ex}} \def\inty{\int_0^\iy} \def\F{\mathcal{F}}

\hfill September 25, 2016

\bc{\large\bf On the Ground State Energy of the Delta-Function\\\vspace{1ex}   Fermi Gas II: Further Asymptotics}\ec

\bc{\large\bf Craig A.~Tracy}\\
{\it Department of Mathematics \\
University of California\\
Davis, CA 95616, USA}\ec

\bc{\large \bf Harold Widom}\\
{\it Department of Mathematics\\
University of California\\
Santa Cruz, CA 95064, USA}\ec

\begin{abstract} 
Building on previous work of the authors, we here derive the weak coupling asymptotics to order $\g^2$ of the ground state energy of the delta-function Fermi gas. We use a method that can be applied to a large class of finite convolution operators.

\end{abstract}

\bc{\bf 1. Introduction}\ec

One of the most widely studied Bethe Ansatz solvable models is the quantum, many-body system in one-dimension with   delta-function two-body interaction \cite{LL} with
Hamiltonian
\[ H_N= -\sum_{j=1}^N \frac{\partial^2}{\partial x_j^2} + 2 c \sum_{i<j} \delta(x_i-x_j). \]
Here $N$ is the number of particles and $2c$ is the coupling constant.
   A basic quantity is  the ground state energy per particle in
the thermodynamic limit:  If $E_0(N,L)$ is the ground state energy for the finite system of $N$ particles on a circle of length $L$,  then in the limit $N\rightarrow \infty$, $L\rightarrow\infty$, such
that $\rho:=N/L$ is fixed, the ground state energy per particle is
\[ \varepsilon_0 := \lim \frac{E_0(N,L)}{N}. \] 
Lieb and Liniger \cite{LL}  showed, for particles with Bose statistics and repulsive interaction ($c>0$), that $\fe_B:=\varepsilon_0/\rho^2$ is a function only of $\gamma:= c/\rho$. To
state their result, we define the \textit{Lieb-Liniger operator}
\be \mathcal{L}_\kappa f(x):= \frac{\kappa}{\pi}\int_{-1}^1 \frac{f( y)}{(x-y)^2+\kappa^2}\, dy,\>\>\> -1<x<1. \label{LLop}\ee
If $f_B(x;\k)$ solves the Lieb-Liniger integral equation
\be f(x) -\mathcal{L}_\kappa f(x) =1 ,\label{LLeqn}\ee
then $\fe_B(\gamma)$ is determined, by elimination of $\kappa$,  from the relations
\[ \frac{\kappa}{\gamma}=\frac{1}{2\pi} \int_{-1}^1 f_B(x;\kappa)\,dx, \>\>\> \fe_B(\g) = \frac{1}{2\pi} \left(\frac{\gamma}{\kappa}\right)^3 \int_{-1}^1 x^2 f_B(x;\kappa) \, dx. \]
The asymptotics of $\fe_B(\g)$ as $\g\to0$ have been derived in the literature. See \cite{TW1} and references therein.
 
A natural question  to ask is how the problem changes when the  particles obey Fermi statistics.  The generalization of Bethe Ansatz to this case   was solved  by Gaudin \cite{gaudin, gaudinBook} and Yang \cite{yang}.  For spin-1/2 particles with attractive interaction ($c<0$) with total spin zero, the ground state energy per particle in the thermodynamic limit is given by \cite{gaudinBook}
 \[ \frac{\varepsilon_0}{\rho^2} = -\frac{\gamma^2}{4}+\fe_F(\gamma) \]
 where $\gamma = \vert c\vert/\rho$ and  the equation is now the {\it Gaudin integral equation}
  \be f(x) + \mathcal{L}_\kappa f(x) =1.\label{gaudineqn}\ee
 If $f_F(x;\k)$ solves this equation, 
 then $\fe_F(\g)$ is determined by elimination of $\k$  from the equations
\be \frac{\kappa}{\gamma}=\frac{2}{\pi} \int_{-1}^1 f_F(x;\kappa)\,dx, \>\>\> \fe_F(\gamma) =\frac{2}{\pi} \left(\frac{\gamma}{\kappa}\right)^3 \int_{-1}^1 x^2 f_F(x;\kappa) \, dx. \label{eqns}\ee
 
Equation (\ref{gaudineqn})  also arises in the computation of the charge $Q$
on each
of two coaxial conducting discs of radius one separated by a distance $\kappa$ and each maintained at the same unit potential. For the Lieb-Liniger equation (\ref{LLeqn}), the discs are maintained at equal  but opposite
potentials.  In both cases the charge $Q$ is given by a constant times the zeroth moment of $f$. For the case of equal potentials (the Fermi case), the charge is given by
\be Q={1\ov\pi}\int_{-1}^1 f_F(x;\k)\,dx,\label{Q}\ee
and Leppington and Levine \cite{LepLev} proved rigorously that as $\k\to0$,
\be Q ={1\ov\pi} + {\k\ov2\pi^2}\left(\log\k\inv+\log\pi+1\right)+o(\kappa). \label{Q0}\ee
The authors derived this by finding an approximate solution of the related boundary value problem.  In later work, Atkinson and Leppington \cite{al} analyzed the integral equation directly and reproduced this result.

As for the ground state energy, Gaudin \cite{gaudinBook} used an approximate solution to (\ref{gaudineqn}) to obtain
\be \fe_F(\g) = \frac{\pi^2}{12} - \frac{\gamma}{2} +\textrm{o}(\gamma).\label{eF1}\ee
Guan and Ma \cite{GM} derived (\ref{eF1}) with the error bound $O(\g^2)$, although Krivnov and Ovchinnikov \cite{KO} had predicted earlier that the term $\g^2 \log^2\g^{-1}$ appears. Using different methods to analyze (\ref{gaudineqn}), Iida and Wadati \cite{iw} found 
\be\fe_F(\gamma) = \frac{\pi^2}{12} - \frac{\gamma}{2} +\frac{\gamma^2}{6}+o(\gamma^2).\label{IW}\ee

The methods used to derive the above-mentioned results for $\fe_F(\g)$ were heuristic. In \cite{TW2} we applied a rigorous analysis to the integral equation (\ref{gaudineqn}) to derive the first-order results (\ref{Q0}) and (\ref{eF1}). It was indicated there that one could in principle derive further asymptotics, and this is what we do here. 

We use the notation $O(\k^{n+})$ to denote a bound $O(\k^n\,\log^m\k\inv)$ for some $m\ge0$, and similarly for $O(\g^{n+})$. What we have found is that as $\k\to0$,
\be Q ={1\ov\pi} + {\k\ov2\pi^2}(\log\k\inv+\log\pi+1)+{\k^2\ov4\pi^3}(\log\k\inv+\log\pi+1/2)+O(\k^{3+}),\label{Q2}\ee
and as $\g\to0$,
\be\fe_F(\g) ={\pi^2\ov12}-{\g\ov2}+{\g^2\ov6}+O(\g^{3+}),\label{e0}\ee
thus confirming the Iida-Wadati result (\ref{IW}).

The derivation of these asymptotics involved some straightforward but tedious computations that were done by Maple, and so we cannot claim complete rigor. Until the end we shall present the results of only a few of the preliminary computations; but we shall get to the points where it is clear that those computations were routine. The reader, if he or she so chooses, can check the outcomes that we exhibit.

In the next section we summarize the results of \cite{TW2}. In the following sections we show how to go further.

\bc{\bf 2. Asymptotic solution of the Gaudin equation}\ec

The method used in this section to analyze the Gaudin operator will be seen to be quite general and applicable to a large class of finite convolution operators. It was used earlier by one of the autheors \cite{w} to derive asymptotics for Toeplitz matrices, which are the discrete analogue of convolution operators.

We first replace the operator $\mathcal{L}_\k$ with kernel
\[{\k\ov\pi}{1\ov(x-y)^2+\k^2}\]
on the fixed interval $(-1,1)$ by the operator with kernel
\[{1\ov\pi}{1\ov(x-y)^2+1}\]
on the variable interval $(-1/\k,\,1/\k)$. For convenience we set $r=2/\k$ and consider the convolution equation  
\[{f(x)\ov2}+{1\ov2\pi}\int_{-r/2}^{r/2}{f(y)\ov (x-y)^2+1}\,dy=1,\ \ \ -r/2<x<r/2.\] 
(The factors $1/2$ here avoid factors $\sqrt2$ later.)

The solution $f_F(x;\k)$ of (\ref{gaudineqn}) and our $f(x)$ are related by $f(rx/2)=2\,f_F(x;\k)$. From (\ref{Q}) we get
\be Q={1\ov r\pi}\,\int_{-r/2}^{r/2}f(x)\,dx={\k\ov2\pi}\,\int_{-r/2}^{r/2}f(x)\,dx.\label{Q1}\ee
From the first part of (\ref{eqns}) we find that
\be\g=\({1\ov\pi}\int_{-r/2}^{r/2} f(x)\,dx\)\inv={1\ov2}\k \,Q\inv.\label{gamma}\ee
From (\ref{eqns}) and a little computation we find that
\be\fe_F(\g)=\pi^2{{\displaystyle\int_{-r/2}^{r/2} x^2\,f(x)\,dx\ov\(\displaystyle\int_{-r/2}^{r/2} f(x)\,dx\)^3}}.\label{e}\ee

Now we go to our integral equation. If we extend the function $f(x)$ to be zero outside the interval $(-r/2,r/2)$ then the equation may be written
\[\int_{-\iy}^\iy k(x-y)\,f(y)\,dy=g(x),\ \ \ x\in (-r/2,r/2),\]
where
\[k(x)={1\ov2}\,\d(x)+{1\ov2\pi}{1\ov x^2+1},\ \ \ g(x)=\ch_{(-r/2,r/2)}(x).\]

The Fourier transforms\footnote{In our notation, the $x\to\x$ Fourier transform has $e^{ix\x}$ in the integrand; the $\x\to x$ inverse \Ft has $e^{-ix\x}$ in the integrand and the factor $1/2\pi$} $\s(\x)$ of $k$ and $\gh(\x)$ of $g$ are given by 
\be\s(\x)=(1+e^{-|\x|})/2,\ \ \ \gh(\x)=2\,\sin(r\x/2)/\x.\label{sgFts}\ee
If $\fh$ is the \Ft of $f$  then $\s\fh-\gh$ is the \Ft of an $L^1$ function supported outside the interval $(-r/2,r/2)$. Such a function may be written as $e^{ir\x/2}\,h^+(\x)+e^{-ir\x/2}\,h^-(\x)$, where $h^\pm$ is the \Ft of an $L^1$ function supported on $\R^\pm$. Thus, by taking Fourier transforms we may rewrite the equation as
\[ \s\fh=\gh+e^{ir\x/2}\,h^++e^{-ir\x/2}\,h^-.\]
We consider $h^\pm$ the unknown functions; once they are determined, so is $\fh$.

We denote by $\ps\to \ps_\pm$ the conjugate by the \Ft of multiplication by $\ch_{\R^\pm}$, the charactristic functions of $\R^\pm$. These are given by
\[\ps_\pm(\x)={1\ov2}\ps(\x)\pm{1\ov2\pi i}\,\int\iyy {\ps(\e)\ov \e-\x}\,d\e,\]
where the integral is a principal value. The functions extend analytically to the upper and lower half-planes by the formulas
\be\ps_\pm(\x)=\pm{1\ov2\pi i}\,\int\iyy {\ps(\e)\ov \e-\x}\,d\e,\label{pmoperators}\ee
where $\x$ is in the upper half-plane for $\ps_+$ and the lower half-plane for $\ps_-$. 

The {\it Wiener-Hopf factors} of $\s$, which confusingly we denote by $\s_\pm$, are given by
\[\s_\pm=e^{(\log\s)_\pm},\]
where the $\pm$ on the right are the projection operators defined above. The function $\log\s$ is a constant plus the \Ft of an $L^1$ function. It follows that $\s_\pm$ and their reciprocals are constants plus Fourier transforms of $L^1$ functions supported on $\R^\pm$. It follows that by changing notation we may replace our equation by
\be \s_-\s_+\fh=\gh+e^{ir\x/2}\,\s_+\,h^++e^{-ir\x/2}\,\s_-h^-.\label{eq}\ee
The factors are given explicitly by \cite{al}
\[\s_+(\x)=\pi^{1/2}\,\exp\left\{{\x\ov2\pi i}\Big[\log(-i\x)-\log2\pi-1\Big]\right\}\,\G\({1\ov2}+{\xi\ov2\pi i}\)\inv,\]
\[\s_-(\x)=\pi^{1/2}\,\exp\left\{-{\x\ov2\pi i}\Big[\log(i\x)-\log2\pi-1\Big]\right\}\,\G\({1\ov2}-{\xi\ov2\pi i}\)\inv.\]
For $\x$ in the upper resp. lower half-plane, $-i\x$ resp. $i\x$ lies in the right half-plane and the principal values of the logarithms are taken.

Since $\s_\pm(0)=1$ and $\gh(0)=r$, we have 
\be\int_{-r/2}^{r/2}f(x)\,dx=\fh (0)=r+h^+(0)+h^-(0),\label{rhh}\ee
which by (\ref{Q1}) determines $Q$ from $h^\pm(0)$. 
Observe also that
\[\int\irr x^2\,f(x)\,dx=-\fh''(0)\] 
\be=\fh(0)/2-2\times \textrm{the coefficient of $\x^2$ in the expansion of}\ \s(\x)\fh(\x).\label{moment}\ee
So the goal is to find the coefficients in the expansions of $h^\pm(\x)$ as $\xi\to0$. 

Here is how we do this. The inverse \Ft of $\fh$ is supported on $(-r/2,\iy)$, so the inverse \Ft of $e^{ir\x/2}\s_+\fh$ is supported on $\R^+$. Therefore if we multiply 
(\ref{eq}) by $e^{ir\x/2}/\s_-$ and apply the minus operator we get
\[0=(e^{ir\x/2}\,\s_+\,\fh)_-=\({e^{ir\x/2}\,\gh\ov\s_-}\)_-+\(e^{ir\x}\,{\s_+\ov\s_-}\,h^+\)_-+h^-.\]
Similarly
\[0=\({e^{-ir\x/2}\,\gh\ov\s_+}\)_++h^++\(e^{-ir\x}\,{\s_-\ov\s_+}\,h^-\)_+.\]

Define the operators $U$ and $V$ by 
\be Uu^-=\(e^{-ir\x}\,{\s_-\ov\s_+}\,u^-\)_+,\ \ \ Vv^+=\(e^{ir\x}\,{\s_+\ov\s_-}\,v^+\)_-.\label{UV}\ee
The operator $U$ takes Fourier transforms of functions in $L^1(\R^-)$ to Fourier transforms of functions in $L^1(\R^+)$, and $V$ does the opposite. If we define
\be G^-=-\({e^{ir\x/2}\,\gh\ov\s_-}\)_-,\ \ \ G^+=-\({e^{-ir\x/2}\,\gh\ov\s_+}\)_+,\label{G}\ee
our two relations may be written
\[h^-+Vh^+=G^-,\ \ \ h^++Uh^-=G^+.\]
The solution is given, formally, by
\[\twoone{h^-}{h^+}=\(I+\twotwo{0}{V}{U}{0}\)\inv\,\twoone{G^-}{G^+}\]
\be=\sum_{j=0}^\iy(-1)^j\,\twotwo{0}{V}{U}{0}^j\,\twoone{G^-}{G^+}.
\label{heq}\ee

This is quite general. For the Gaudin equation we have a precise statement. Using a different notation than in \cite{TW2}, we define $\F^+$ to be those families of Fourier transforms of functions in $L^1(\R^+)$, depending on the parameter $r$, for which there is an asymptotic expansion as $\x\to0$,
\[\ph(\x)\sim \sum_{0\le m\le n}c_{n,m,r}\,\x^n\,\log^m(-i\x),\]
where each $c_{n,m,r}=O(r^{n+})$ as $r\to\iy$.
Similarly we define $\F^-$.  
It was shown in \cite{TW2} that $G^\pm$ and $h^\pm$ belong to $\F^\pm$, and that truncating the series in (\ref{heq}) at $j=k-1$ leads to an error in $h^\pm$ belonging to $r^{-k}\F^\pm$. (In other words the error equals $r^{-k}$ times a function in $\F^\pm$.)

Taking $k=1$ leads to the conclusion that if in (\ref{eq}) we replace $h^\pm$ on the right side by $G^\pm$ the error in the constant term is $O(r^{-1+})$ and the error in the coefficient of $\x^2$ is $O(r^{1+})$. This implies\footnote{We use that $\F^\pm$ is closed under multiplication by $e^{\pm ir\x/2}$ or $\s_\pm(\x)$.} that the error in $f(0)$ is $O(r^{-1+})$ and the error in $f''(0)$ is $O(r^{1+})$. Having computed the coefficients in the expansions of $G^\pm$ to the right order, this led us to the first-order asymptotics (\ref{Q0}) and (\ref{eF1}).

Here we truncate the series at $j=1$, in other words we make the replacements
\[h^-\to G^--VG^+,\ \ \ h^+\to G^+-UG^-,\]
knowing that this will lead to an error $O(r^{-2+})$ in $f(0)$ and an error $O(r^{0+})$ in $f''(0)$,\footnote{We refer to these as ``acceptable errors''.} which will give the next-order asymptotics of $Q$ and of $\fe_F(\g)$.

\bc{\bf 3. Expansion of \boldmath$G^+(\x)$ near \boldmath$\x=0$}\ec

We do this differently than in \cite{TW2}. From (\ref{sgFts}), (\ref{pmoperators}), and (\ref{G}), we see that the expression for $G^+(\x)$ for $\x$ in the upper half-plane is
\[G^+(\x)={1\ov2\pi}\int\iyy{1-e^{-ir\e}\ov\e\,\s_+(\e)}\,{d\e\ov\e-\x}.\]
Using $\s_+(0)=1$ we write this as the difference
\[{1\ov2\pi}\int\iyy{1\ov\e}\left[{1\ov\s_+(\e)}-1\right]{d\e\ov\e-\x}-
{1\ov2\pi}\int\iyy{1\ov\e}\left[{e^{-ir\e}\ov\s_+(\e)}-1\right]{d\e\ov\e-\x}.\]
For the first integral we push the contour up. We get contributions from the pole at $\e=\x$, with the result
\[{i\ov \x}\left({1\ov \s_+(\x)}-1\right).\]

For the second integral we use a trick from \cite{TW2}. We swing the $\R^+$ part of the contour down to the right side of the negative imaginary axis and the $\R^-$ part of the contour down to the left side of the negative imaginary axis, making there the substitution $\e=-ix$. We use
$1/\s_+=\s_-/\s$ and that the analytic continuation of $1/\s(\e)$ to the right side of the imaginary axis minus its analytic continuation to the left side of the imaginary axis equals 
\[{2\ov 1+e^{ix}}-{2\ov 1+e^{-ix}}=-2i\tan(x/2).\]
Thus the second integral with its factor becomes 
\[\inty{e^{-rx}\ov x}\,\ps(x)\,{dx\ov x-i\x},\ \ \ \ \ps(x)={1\ov\pi}\,\s_-(-ix)\,\tan(x/2).\]
(This is a principal value integral at each odd multiple of $\pi$. The contributions of the integrals over the little semicircles on either side of the imaginatry axis cancel each other.) Observe that with $\x$ in the upper half-plane $-i\x$ is in the right half-plane.

We have obtained the representation
\be G^+(\x)={i\ov \x}\left({1\ov \s_+(\x)}-1\right)-\inty{e^{-rx}\ov x}\,\ps(x)\,{dx\ov x-i\x}.\label{Grep}\ee
We are interested first in the first few terms in the expansion of this as $\x\to0$, and for $k\le2$ we allow an error $O(r^{k-2+})$ in the coefficient of terms involving $\x^k$.\footnote{Recall that the expansion of $G^+(\x)$ involves powers of $\x$ times powers of logarithms. Powers of logarithms also occur in the expansion of $\ps(x)$,} The expansion of the first summand can be found and has terms independent of $r$. For the integral, if we replace $\ps(x)$ by the terms up to powers less than $N$ of its expansion near $x=0$ the error will be an integral in which $\ps(x)$ is replaced by $O(x^{N+})$.\footnote{The reader may be concerned about the factor $\tan x/2$ in $\ps(x)$. A representation using a somewhat different contour resolves this issue. We deform the original contour to the left and right parts of the negative imaginary axis only down to $-i\pi/2$, say, and then slightly off-vertical rays from $-i\pi/2$ downward. There are no singularities on the modified contour. The integrals over the rays are analytic in $\x$ near zero with the coefficient of each power of $\x$ exponentially small in $r$. The upper limit on the integral in (\ref{Grep}) becomes $\pi/2$, which does not affect the argument that follows.\label{tan}} In the $\x$-expansion of the resulting integral the coefficients involving $\x^k$ would be $O(r^{-N+k+1+})$. This shows that with acceptable errors we allow in the coefficients we may replace $\ps(x)$ by the terms in its expansion up to powers less than three, and these are 
\[{x\ov2\pi}-{x^2\ov4\pi^2}\,(\log x\inv+\log(\pi/2)-\g_E+1).\]
(Here and below $\g_E$ denotes the Euler gamma.)
From the general formula
\be\int_0^\iy e^{-x}\,x^{a-1}\,{dx\ov x+z}=\G(a)\,e^z\,E_a(z),\label{Ei}\ee
we see that the integrals that arise can be expressed in terms of generalized exponential integrals (and a derivative of one of them) evaluated at $-ir\x$, and their expansions are known. 

\bc{\bf 4. Expansion of \boldmath$VG^+(\x)$ near \boldmath$\x=0$}\ec

From (\ref{pmoperators}) and the definition of the operator $V$ in (\ref{UV}) 
we have
\[VG^+(\x)=-{1\ov2\pi i}\int_{-\iy}^\iy e^{ir\e}\,{\s_+(\e)\ov\s_-(\e)}\,G^+(\e)\,{d\e\ov\e-\x},\]
where $\x$ is now in the lower half-plane. We use the same trick as in the last section. We rewrite $\s_+/\s_-$ as $\s_+^2/\s$ and swing the half-lines up to the imaginary axis in the upper half-plane, where we make the substitution $\e=i y$. The result is 
\[VG^+(\x)=\inty e^{-ry}\,\ph(y)\,G^+(iy)\,{dy\ov y+i\x},\ \ \ \ph(y)=-{1\ov\pi}\s_+(iy)^2\,\tan(y/2).\]
Now $i\x$ is in the right half-plane. 

In formula (\ref{Grep}) $G^+(\x)$ is given as the sum of two terms, the first of which is independeent of $r$. Its contribution to the integral for $VG^+(\x)$ is like the integral in (\ref{Grep}) and we can treat it analogously. In this case for the error we accept in the coefficients we may replace $\ph(y)$ times the first term in (\ref{Grep}) (with $\x$ replaced by $iy$) by the terms in its expansion up to powers less than two. Thus we may replace this product by the single term
\[-{1\ov4\pi^2}(\log y\inv+\log(\pi/2)-\g_E+1)\,y.\]
Again the corresponding integral over $y$ is expressible in terms of  exponential integrals.

There remains the double integral
\[\inty e^{-ry}\,\ph(y)\,{dy\ov y+i\x}\,\inty{e^{-rx}\ov x}\,\ps(x)\,{dx\ov x+y}.\]
As in footnote \ref{tan} we may replace the upper limits of integration by $\pi/2$, so the integrands have no singularities. Now, if replace $\ps(x)$ in the inner integral by the terms up to powers less than $N$ of its expansion near $x=0$ the error in the inner integral can be seen to be $O(\min(y^{N-1+},r^{-N+1+}))$. The resulting double integral would be a function of $\x$ for whose expansion the coefficient of $\x^k$ (for fixed $k<N$) would be $O(r^{-N+k+})$. So given our acceptable error we may replace $\ps(x)$ by finitely many terms in its expansion. Then we may replace $\ph(y)$ by finitely many terms in its expansion. (And we may replace the upper limits by $\iy$.) 

To see what the individual summands will be we consider the integral
\[\inty e^{-ry}\,y^p\,{dy\ov y+i\x}\;\inty e^{-rx}\,x^q\,{dx\ov x+y}\] 
with $p\ge1,\ q\ge0$. (When logarithms appear we differentiate some number of times with respect to $p$ or $q$.)
We make the variable changes $y\to y/r,\ x\to x/r$ and set $X=ir\x$. We obtain, using (\ref{Ei}),
\[r^{-p-q}\,\inty e^{-y}\,y^p\,{dy\ov y+X}\;\inty e^{-x}\,x^q\,{dx\ov x+y}=r^{-p-q}\,\G(q+1)\,\inty y^p\,E_{q+1}(y){dy\ov y+X}.\]
From the expansion of $E_{q+1}(y)$ as $y\to0$ we see that there is an expansion as $X\to0$ with summands that are nonnegative powers times logarithms. In terms of $\x$ (recall that $X=ir\x$), the summands involving $\x^k$ have coefficients $O(r^{k+})$. Recalling that $r^{-p-q}$ multiplies this integral, and the errors in the coefficients that are acceptable, we see that we need consider only the term with $p=1,\ q=0$. The integral, which eventually gets the factor $-1/(4\pi^2 r)$, becomes
\[\inty y\,E_1(y)\,{dy\ov y+X}=X\,\inty E_1(Xy)\,{y\,dy\ov y+1}.\]
Integrating by parts using $E_1'(y)=-y\inv\,e^{-y}$ gives
\be 1-X\inty e^{-Xy}\,y\inv\,\log(y+1)\,dy.\label{E1}\ee

The following may be a needlessly complicated way of finding the asymptotics of this as $X\to0$.. If we call the last integral $I(X)$ then
\[I'(X)=-\inty e^{-X\,y}\log(1+y)\,dy={d\ov ds}\inty e^{-X\,y}(1+y)^{-s}\,dy\Big|_{s=0}=e^X\,{d\ov ds}E_s(X)\Big|_{s=0}.\]
From the known expansion of $E_s(X)$ we find that as $X\to0$,
\[I'(X)=X\inv(\log X+\g_E)+\log X-1+\g_E+O(X^{1+}).\]
Integrating from 1 to $X$ gives
\[I(X)={1\ov2}\log^2 X+\g_E\,\log X+C+X\,\log X-(2-\g_E)X+O(X^{2+}),\]
for some constant $C$. To evaluate $C$ we use
\[-X\inv(\log X+\g_E)=\inty e^{-Xy}\,\log y\,dy,\]
and integrate from $X$ to 1 to obtain
\[{1\ov2}\log^2 X+\g_E\log X=\inty {e^{-Xy}-e^{-y}\ov y}\,\log y\,dy.\]
Subtracting this from $I(X)$ and taking the $X\to0$ limit gives
\[C=\inty [\log(1+y)-(1-e^{-y})\log y]\,{dy\ov y}.\]
We leave as an exercise for the reader that $C=\pi^2/4+\g_E^2/2$.
Thus (\ref{E1}), which gets the factor $-1/(4\pi^2r)$, has the asymptotics as $X\to0$,
\[1-[(\log^2X)/2 +\g_E\,\log X+\pi^2/4+\g_E^2/2]\,X-[\log X-2+\g_E]\,X^2+O(X^{3+}).\]
After setting $X=ir\x$ this gives the asymptotics as $\x\to0$, with terms up to those involving $\x^2$, and with acceptable error for the coefficients.

\bc{\bf 5. The final results}\ec

We know that with acceptable error we may replace $h^-(\x)$ in the right side of (\ref{eq}) by $G^-(\x)-VG^+(\x)$. In section 3 we showed how to compute the series for $G^+(\x)$ up to terms involving $\x^2$, and $G^-(\x)$ is the complex conjugate of $G^+(\x)$ for real $\x$. In  section~4 we showed how to compute the series for $VG^+(\x)$ up to terms involving $\x^2$. All with acceptable errors in the coefficients. Then we multiply by $e^{-ir\x/2}\,\s_-(\x)$ to obtain the last term on the right in (\ref{eq}). The next-to-last term is the complex conjugate of the last, so we know that, also. Then we add the series for $\gh(\x)$. From these computations and (\ref{rhh}) we find that
\[\int_{-r/2}^{r/2}f(x)\,dx=r+{1\ov\pi}\,[\log(\pi r/2)+1]+{1\ov\pi^2r}\,[\log(\pi r/2)+1/2]+O(r^{-2+}).\]
(Observe that all terms involving $\g_E$ have cancelled.) Setting $r=2/\k$ and using (\ref{Q1}), we obtain (\ref{Q2}).

{}From the computations and (\ref{moment}) we find that
\[\int\irr x^2\,f(x)\,dx={r^3\ov12}+{r^2\ov 4\pi}\,[\log(\pi r/2)-1]+{r\ov4\pi^2}\,[\log^2(\pi r/2)-\log(\pi r/2)-5/2+2\pi^2/3]+O(r^{0+}).\]
Then from these and (\ref{e}) we obtain 
\[\fe_F(\g)={\pi^2\ov12}-{\pi\ov2r}+{1\ov 2r^2}\,[\log(\pi r/2)+1+\pi^2/3]+O(r^{-3+}).\]
From (\ref{gamma}) we find that
\[\g={\pi\ov r}-{1\ov r^2}\,[\log(\pi r/2)+1]+O(r^{-3+}),\]
and (\ref{e0}) follows.

\bc{\bf Acknowledgments}\ec

This work was supported by the National Science Foundation through grants DMS--1207995 (first author) and DMS--1400248 (second author).

\end{document}